\documentstyle[11pt]{article}

\newcommand{\be}{\begin{equation}}
\newcommand{\ee}{\end{equation}}
\newcommand{\bi}[1]{\vspace{-3mm} \bibitem{#1}}

\begin{document}
{\it MODERN PHYSICS LETTERS B, Vol.17. No.23. (2003) 1219-1226.}\\

\begin{center}
{\Large \bf Classical Canonical Distribution for Dissipative Systems}
\vskip 5 mm
{\Large \bf Vasily E. Tarasov } \\

\vskip 3mm
{\it Skobeltsyn Institute of Nuclear Physics,
Moscow State University, Moscow 119992, Russia}

{E-mail: tarasov@theory.sinp.msu.ru}
\end{center}

\begin{abstract}
In this paper we derive the canonical distribution
as a stationary solution of the Liouville equation for
the classical dissipative system.
Dissipative classical systems can have stationary states
look like canonical Gibbs distributions.
The condition for non-potential forces which leads to 
this stationary solution is very simple:
the power of the non-potential forces must be directly proportional
to the velocity of the Gibbs phase (phase entropy density) change.
The example of the canonical distribution for a linear oscillator
with friction is considered.
\end{abstract}

PACS {05.20.-y; 05.30.-d}


\section{Introduction}

The canonical distribution was defined 101 years ago in
the book "Elementary principles in statistical mechanics,
developed with especial reference to the rational foundation
of thermodynamics" \cite{Gibbs} published in 1902.
The canonical distribution function usually
can be derived as a stationary solution of the Liouville
equation for non-dissipative Hamiltonian N-particle systems
with a special set of potential forces \cite{GS1}-\cite{Petrina}.

In general, classical systems are not Hamiltonian systems
and the forces which act on particles are the sum of potential
and non-potential forces.
The non-potential internal forces for an N-particle system
can be connected with nonelastic collisions \cite{Sandu}.
Dissipative and non-Hamiltonian systems can have the same
stationary states as Hamiltonian systems \cite{Tarpla2}.
For example, dissipative quantum systems have pure
stationary states of linear harmonic oscillator \cite{Tarpre}.
We can assume that canonical distribution exists
for classical dissipative systems.

In this paper we consider the Liouville equation for
classical dissipative and non-Hamiltonian N-particle
systems. This equation is the equation of continuity
in 6N dimensional phase space. We find the condition
for the non-potential forces which leads to the stationary
solution of this equation look like the canonical 
distribution function. This condition is very simple:
{\it the power of non-potential forces must be directly
proportional to the velocity of the Gibbs phase
(phase entropy density) change.} Note that the
velocity of the phase entropy density change
is equal to the velocity of the phase volume change.

In Sections 2, the mathematical background and notations are considered.
In this section, we formulate the main conditions for non-potential
forces and derive the canonical distribution from an N-particle
Liouville equation in Hamilton picture. In the Section 3,
in the Liouville picture we substitute the canonical
distribution function in the Liouville equation for a dissipative system
and derive condition for non-potential forces.
In this section, we consider the Maxwell-Boltzmann distribution
function for dissipative systems.
In Section 4, the example of the canonical distribution for linear
oscillator with friction is considered.
Finally, a short conclusion is given in Section 5.

\section{Canonical distribution from Liouville equation
in Hamilton picture}

Let us consider the N-particle classical system in the Hamilton picture.
In general, the equation of motion the $i$th particle,
where $i=1,...,N$, has the form
\[ \frac{d\vec{r}_i}{dt}=\frac{\vec{p}_i}{m},\quad
\frac{d\vec{p}_i}{dt}=\vec{F}_i,\]
where $\vec{F}_i$ is a resulting force, which acts on the $i$th particle.
In the general case, the force is not a potential and we can write
\be \label{F} \vec{F}_i=-\frac{\partial U}{\partial \vec{r}_i}
+\vec{F}^{(n)}_i,\ee
where $U=U(\vec{r})$  is the potential energy of the system,
$\vec{F}^{(n)}_i$ is the sum of non-potential forces
(internal and external), which act on the $i$th particle.
For any classical observable $A=A(\vec{r}_{t},\vec{p}_{t},t)$, where
$\vec{r}=(\vec{r}_1,...,\vec{r}_N)$ and
$\vec{p}=(\vec{p}_1,...,\vec{p}_N)$ in the Hamilton picture,
we have
\be \label{dA} \frac{dA}{dt}=\frac{\partial A}{\partial t}+
\sum^N_{i=1}\frac{\vec{p}_i}{m}
\frac{\partial A}{\partial \vec{r}_i}+
\sum^N_{i=1}\vec{F}_i \frac{\partial A}{\partial \vec{p}_i} .\ee
The Hamiltonian of this system
\be \label{H} H(\vec{r},\vec{p})=
\sum^N_{i=1}\frac{p^2_i}{2m}+U(r) \ee
is not a constant along the trajectory in the 6N dimensional phase space.
From equations (\ref{dA}) and (\ref{F}) we have
\[ \frac{dH}{dt}=
\sum^N_{i=1}\frac{\vec{p}_i}{m}
\frac{\partial U}{\partial \vec{r}_i}+
\sum^N_{i=1}\Bigl(-\frac{\partial U}{\partial \vec{r}_i}
+\vec{F}^{(n)}_i \Bigr) \frac{\vec{p}_i}{m},\]
i.e.
\be \label{dH} \frac{dH}{dt}=
\sum^N_{i=1}(\vec{F}^{(n)}_i,\vec{v}_i ), \ee
where $\vec{v}_i=\vec{p}_i/m$. Therefore, the energy change
is equal to a power ${\cal P}$
of the 
non-potential forces $\vec{F}^{(n)}_i$:
\be \label{P} {\cal P}(\vec{r},\vec{p},t)=
\sum^N_{i=1}(\vec{F}^{(n)}_i,\vec{v}_i ) . \ee

The N-particle distribution function in the Hamilton picture
is normalized using
\be \label{f=1} \int \rho_N(\vec{r}_{t},\vec{p}_{t},t)
d^N\vec{r}_{t} d^N\vec{p}_{t}=1. \ee
The evolution equation of the function
$\rho_N(\vec{r}_{t},\vec{p}_{t},t)$ is the Liouville equation
in the Hamilton picture (for the Euler variables) which has the form
\be \label{Liu1} \frac{d\rho_N(\vec{r}_{t},\vec{p}_{t},t)}{dt}=
-\Omega (\vec{r}_{t},\vec{p}_{t},t)\rho_N(\vec{r}_{t},\vec{p}_{t},t). \ee
This equation describes the change of distribution function
$\rho_N$ along the trajectory in 6N-dimensional phase space.
Here, $\Omega$ is defined by
\be \label{O} \Omega(\vec{r},\vec{p},t)=
\sum^N_{i=1}\frac{\partial \vec{F}^{(n)}_i}{\partial \vec{p}_i},\ee
and $d/dt$ is a total time derivative (\ref{dA}):
\[ \frac{d}{dt}=\frac{\partial}{\partial t}+
\sum^N_{i=1}\frac{\vec{p}_i}{m} \frac{\partial}{\partial \vec{r}_i}+
\sum^N_{i=1}\vec{F}_i \frac{\partial}{\partial \vec{p}_i} . \]
If $\Omega<0$, then the system is called a dissipative system.
If $\Omega \not=0$, then the system is a generalized
dissipative system.
In the Liouville picture the function $\Omega$ is
equal to the velocity of the phase volume change \cite{Tarkn1}.

Let us define a phase density of entropy by
\[ S(\vec{r}_{t},\vec{p}_{t},t)=
-k \ ln \ \rho_N(\vec{r}_{t},\vec{p}_{t},t).\]
This function usually called a Gibbs phase.
Equation (\ref{Liu1}) leads to the equation
for the Gibbs phase:
\be \label{dS} \frac{dS(\vec{r}_{t},\vec{p}_{t},t)}{dt}=
k\Omega (\vec{r}_{t},\vec{p}_{t},t) .\ee
Therefore, the function $\Omega$ is proportional to the
velocity of the phase entropy density (Gibbs phase) change.


Let us assume that the power ${\cal P}(\vec{r}_{t},\vec{p}_{t},t)$
of the non-potential forces
is
directly proportional to the velocity of
the Gibbs phase (phase density of entropy)
change \ $\Omega(\vec{r}_{t},\vec{p}_{t},t)$:
\be \label{PO} {\cal P}(\vec{r}_{t},\vec{p}_{t},t)
=kT\Omega(\vec{r}_{t},\vec{p}_{t},t), \ee
with some coefficient $T$ which is not depend
on $(\vec{r}_{t},\vec{p}_{t},t)$, i.e.  $dT/dt=0$.

Using equations (\ref{dH},\ref{P}) and (\ref{dS}), assumption
(\ref{PO}) can be rewtitten in the form
\[ \frac{dH(\vec{r}_{t},\vec{p}_{t})}{dt}=
T \frac{dS(\vec{r}_{t},\vec{p}_{t},t)}{dt}. \]
Since coefficient $T$ is constant,  we have
\[ \frac{d}{dt}\Bigl( H(\vec{r}_{t},\vec{p}_{t})-
T S(\vec{r}_{t},\vec{p}_{t},t) \Bigr)=0, \]
i.e. the value $(H-TS)$ is a constant along the trajectory
of the system in 6N-dimensional phase space.
Let us denote this constant value by ${\cal F}$.
Then we have
\[ H(\vec{r}_{t},\vec{p}_{t})-
T S(\vec{r}_{t},\vec{p}_{t},t)={\cal F},\]
where $d{\cal F}/dt=0$, i.e.
\[ ln \ \rho_N(\vec{r}_{t},\vec{p}_{t},t)=\frac{1}{kT}\Bigl(
{\cal F}- H(\vec{r}_{t},\vec{p}_{t}) \Bigr) . \]
As the result we have a canonical distribution function
\[ \rho_N(\vec{r}_{t},\vec{p}_{t},t)=exp \frac{1}{kT}\Bigl(
{\cal F}- H(\vec{r}_{t},\vec{p}_{t}) \Bigr) \]
in the Hamilton picture. The value ${\cal F}$ is defined
by normalization condition (\ref{f=1}).

Note that $N$ is an arbitrary natural number since we do not
use the condition $N>>1$ or $N \rightarrow \infty$.

\section{Canonical distribution  in Liouville picture}

Let us consider the Liouville equation for the N-particle
distribution function $\rho_N=\rho_{N}(\vec{r},\vec{p},t)$
in the Liouville picture (for the Lagnangian variables):
\be \label{fN} \frac{\partial \rho_N}{\partial t}+
\sum^N_{i=1}
\frac{\vec{p}_i}{m} \frac{\partial \rho_N}{\partial \vec{r}_i}+
\sum^N_{i=1}
\frac{\partial}{\partial \vec{p}_i} \Bigl(\vec{F}_i \rho_N\Bigr)=0 .\ee
In general, the forces $\vec{F}_i$ are non-potential forces.
This equation is the equation of continuity for 
6N-dimensional phase space. Substituting the canonical
distribution function
\[ \rho_N(\vec{r},\vec{p},t)=exp \frac{1}{kT}\Bigl(
{\cal F}- H(\vec{r},\vec{p},t) \Bigr) \]
in equation (\ref{fN}), we get
\[ -\frac{1}{kT}\Bigl(\frac{\partial H}{\partial t}+
\sum^N_{i=1}\frac{\vec{p}_i}{m} \frac{\partial H}{\partial \vec{r}_i}+
\sum^N_{i=1}\vec{F}_i \frac{\partial H}{\partial \vec{p}_i} \Bigr)\rho_N+
\sum^N_{i=1}\frac{\partial \vec{F}_i }{\partial \vec{p}_i} \rho_N=0 .\]
Since $\rho_N$ is not equal to zero, we have
\[ \frac{\partial H}{\partial t}+
\sum^N_{i=1}\frac{\vec{p}_i}{m} \frac{\partial H}{\partial \vec{r}_i}+
\sum^N_{i=1}\vec{F}_i \frac{\partial H}{\partial \vec{p}_i} =kT
\sum^N_{i=1}\frac{\partial \vec{F}_i }{\partial \vec{p}_i} .\]
If the Hamiltonian $H$ has the form (\ref{H}), than
this equation leads to
\[ \sum^N_{i=1}
\frac{\vec{p}_i}{m} \Bigl(\frac{\partial U}{\partial \vec{r}_i}
+\vec{F}_i \Bigr) =kT
\sum^N_{i=1}\frac{\partial \vec{F}_i }{\partial \vec{p}_i} .\]
Substituting equation (\ref{F}) in this equation, we get the following
condition for non-potential forces $F^{(n)}_i$:
\[ \sum^N_{i=1}\Bigl(\frac{\vec{p}_i}{m}, \vec{F}^{(n)}_i\Bigr) =kT
\sum^N_{i=1}\frac{\partial \vec{F}^{(n)}_i }{\partial \vec{p}_i} .\]
Using notations (\ref{P}) and (\ref{O}), we can rewritte
this condition in the form
\[ {\cal P}(\vec{r},\vec{p},t)=kT\Omega(\vec{r},\vec{p},t). \]
As a result we have that the canonical distribution function
is a solution of the Liouville equation for dissipative and
non-Hamiltonian systems if the power of the non-potential forces
is proportional to the velocity of the phase volume change.


Let us consider a chain of Bogoliubov equations
\cite{Bog,Petrina} for the Liouville equation of the
dissipative systems (\ref{fN}) in approximation
\be \label{approx} \rho_2(\vec{r}_1,\vec{p}_1,\vec{r}_2,\vec{p}_2,t)=
\rho_1(\vec{r}_1,\vec{p}_1,t) \rho_1(\vec{r}_2,\vec{p}_2,t) .\ee
The non-potential forces $\vec{F}^{(n)}_i$ in equation (\ref{F}) 
is a sum of external forces ${\vec F}^{(n,e)}_i$
and internal forces $\vec{F}^{(n,i)}_i$.
For example, in the case of binary interactions we have
\[ \vec{F}^{(n)}_i=\vec{F}^{(n,e)}_{i}(\vec{r}_i,\vec{p}_i,t)+
\sum^N_{j=1,j\not=i}
\vec{F}^{(n,i)}_{ij}(\vec{r}_i,\vec{p}_i,\vec{r}_j,\vec{p}_j,t).\]
In approximation (\ref{approx}) we can define the force
\[ \vec{F}_1=-\frac{\partial (U+U_{eff})}{\partial \vec{r}_1}
+\vec{F}^{(n)}_1+\vec{F}^{(n)}_{1,eff}, \]
where
\[ \vec{F}^{(n)}_{1,eff}(\vec{r}_1,\vec{p}_1,t)= 
\int d\vec{r}_2 d\vec{p}_2
\rho_1(\vec{r}_2,\vec{p}_2,t)
\vec{F}^{(n,i)}_{12}(\vec{r}_1,\vec{p}_1,\vec{r}_2,\vec{p}_2,t), \]
\[ U_{eff}(\vec{r}_1,\vec{p}_1,t)=\int d\vec{r}_2 d\vec{p}_2
\rho_1(\vec{r}_2,\vec{p}_2,t) U(|\vec{r}_2-\vec{r}_1|). \]
If we consider the 1-particle distribution then Liouville equation
(\ref{fN}) in approximation (\ref{approx}) has the form
\be \label{f1} \frac{\partial \rho_1}{\partial t}+
\frac{\vec{p}_1}{m} \frac{\partial \rho_1}{\partial \vec{r}_1}+
\frac{\partial}{\partial \vec{p}_1} \Bigl(\vec{F}_1 \rho_1\Bigr)=0 , \ee
where
$ \rho_{1}=\rho_{1}(\vec{r}_{1},\vec{p}_{1},t)$.
Let us consider a condition for the non-potential forces
\[ (\vec{p}_1,\vec{F}^{(n)}_1+\vec{F}^{(n)}_{1,eff})=mkT
\frac{\partial (\vec{F}^{(n)}_1+
\vec{F}^{(n)}_{1,eff})}{\partial \vec{p_1}}.\]
In this case, we can derive the 1-particle
distribution function (as in Section 3) in the form
\[ \rho_1(\vec{r},\vec{p},t)=A \
exp -\frac{1}{kT}\Bigl(\frac{\vec{p}^2}{2m}+U(\vec{r})+
U_{eff}(\vec{r})\Bigr). \]
This is a Maxwell-Bolztmann distribution function.

\section{Canonical distribution for harmonic oscillator with friction}

Let us consider the N-particle system with a linear friction
defined by non-potential forces
\be \label{fric} \vec{F}^{(n)}_i=-\gamma \vec{p}_i, \ee
where $i=1,...,N$. 
Note that $N$ is an arbitrary natural number.
Substituting equation (\ref{fric}) into equations (\ref{P}) 
and (\ref{O}), we get
the power ${\cal P}$ and the Gibbs phase $\Omega$:
\[ {\cal P}=-\frac{\gamma}{m} \sum^N_{i=1} p^2_i, \quad \Omega=-\gamma. \]
Condition (\ref{PO}) has the form
\be \label{PO2} \sum^N_{i=1}\frac{p^{2}_i}{m}=kT, \ee
i.e. the kinetic energy of the system must be a constant.
Note that equation (\ref{PO2}) has no friction
parameter $\gamma$. Condition (\ref{PO2}) is a non-holonomic
(non-integrable) constraint \cite{Dob}.

Let us consider the N-particle system with friction (\ref{fric})
and non-holonomic constraint (\ref{con}).
The equations of motion for this system have the form
\be \label{em} \frac{d\vec{r}_i}{dt}=\frac{\vec{p}_i}{m},\quad
\frac{d\vec{p}_i}{dt}=-\gamma \vec{p}_i-
\frac{\partial U}{\partial \vec{r}_i}+
\lambda \frac{\partial G}{\partial \vec{p}_i}, \ee
where the function $G$ is defined by
\be \label{con} G(\vec{r},\vec{p})=
\frac{1}{2}\Bigl(\sum^N_{i=1}p^{2}_i-mkT \Bigr):
\quad G(\vec{r},\vec{p})=0. \ee
Equations (\ref{em}) with condition (\ref{con})
define 6N+1 variables
$(\vec{r},\vec{p},\lambda)$.

Let us find the Lagrange multiplier $\lambda$.
Substituting equation (\ref{con}) into equation (\ref{em}), we get
\be \label{em2}
\frac{d\vec{p}_i}{dt}=-(\gamma-\lambda) \vec{p}_i-
\frac{\partial U}{\partial \vec{r}_i} . \ee
Multiplying both sides of equation (\ref{em2}) by
$\vec{p}_i/m$ and summing over index $i$, we obtain
\be \label{em3}
\frac{d}{dt}\Bigl(\sum^N_{i=1}\frac{\vec{p}^{2}_i}{2m}\Bigr)=
-(\gamma-\lambda)\sum^N_{i=1} \frac{\vec{p}^{2}_i}{m} -
\sum^N_{i=1}\Bigl(\frac{\vec{p}_i}{m},\frac{\partial U}{\partial \vec{r}_i}\Bigr) . \ee
Using $dG/dt=0$ and substituting  equation (\ref{PO2}) into 
equation (\ref{em3}), we get
\[ 0=-(\gamma-\lambda) kT -
\sum^N_{j=1} (\frac{\vec{p}_j}{m},\frac{\partial U}{\partial \vec{r}_j}) . \]
Therefore, the Lagrange multiplier $\lambda$ is equal to
\[ \lambda= \frac{1}{mkT}
\sum^N_{j=1}\Bigl(\vec{p}_j,\frac{\partial U}{\partial \vec{r}_j}\Bigr)+\gamma . \]
As the result, we have the holonomic system which is equivalent
to the non-holonomic system (\ref{em},\ref{con}) and defined by
\be \label{em4} \frac{d\vec{r}_i}{dt}=\frac{\vec{p}_i}{m},\quad
\frac{d\vec{p}_i}{dt}= \frac{1}{mkT}
\sum^N_{j=1}\Bigl(\vec{p}_j,\frac{\partial U}{\partial \vec{r}_j}\Bigr)
\vec{p}_i-\frac{\partial U}{\partial \vec{r}_i}. \ee
Condition (\ref{PO}) or (\ref{PO2}) for the 
classical N-particle system (\ref{em4}) is satisfied.
If the time evolution of the N-particle system (\ref{em})
has non-holonomic constraints (\ref{con}) or the evolution is defined
by equation (\ref{em4}), then we have the canonical distribution 
function in the form
\[ \rho(\vec{r},\vec{p})=exp\frac{1}{kT}
\Bigl({\cal F}-\sum^N_{i=1}\frac{p^2_i}{2m}-U\Bigr). \]
For example, the N-particle system with the forces
\[ \vec{F}_i= \frac{\omega^2}{kT} \vec{p_i}
\sum^N_{j=1}(\vec{p}_j,\vec{r}_j)-m\omega^2 \vec{r}_i \]
can have a canonical distribution look likes
the canonical distribution of the linear harmonic oscillator:
\[ \rho(\vec{r},\vec{p})=exp\frac{1}{kT}
\Bigl({\cal F}-H(\vec{r},\vec{p})\Bigr), \]
where
\[ H(\vec{r},\vec{p})=\sum^N_{i=1}\frac{p^2_i}{2m}+
\sum^N_{i=1}\frac{m\omega^2 r^2_i}{2}. \]

\section{Conclusion}

Dissipative and non-Hamiltonian classical systems can have
stationary states that look like canonical distribution.
The condition for non-potential forces which leads to
the canonical distribution function for dissipative systems
is very simple: {\it thepower of all non-potential forces must
be directly proportional to the  velocity of the Gibbs phase
(phase entropy density) change.}

In the papers \cite{Tarpla1,Tarmsu}, the quantization
of evolution equations for dissipative and non-Hamiltonian systems
was suggested.
Using this quantization it is easy to derive quantum
Liouville-von Neumann equations for the N-particle matrix density
operator of the dissipative quantum system.
The condition which leads to the canonical matrix density
solution of the Liouville-von Neumann equation can be generalized
for the quantum case by the quantization method
suggested in \cite{Tarpla1,Tarmsu}.


This work was partially supported by the RFBR grant No. 02-02-16444



\end{document}